\documentclass[amsmath,amssymb,floatfix,showpacs,twocolumn]{revtex4-1}
\usepackage{graphicx,amsmath,amssymb,amsfonts}

\begin{document}

\title{Laser control of complete vibrational transfer in Na$_2$ using resonance coalescence}

\author{O. Atabek$^1$, R. Lefebvre$^{1,2}$, M. Lepers$^3$, A. Jaouadi$^{1,3}$, O. Dulieu$^3$, and V. Kokoouline$^{3,4}$}
\affiliation{$^1$Institut des Sciences Mol\'eculaires d'Orsay, B\^at 350, Univ. Paris-Sud (CNRS),  91405 Orsay France\\
$^2$U.F.R. de Physique Fondamentale et Appliqu\'ee, Universit\'e Pierre et Marie Curie, 75321 Paris, France\\
$^3$Laboratoire Aim\'e Cotton, CNRS,  B\^at. 505, Univ. Paris-Sud, 91405 Orsay, France\\
$^4$Department of Physics, University of Central Florida, Orlando, FL 32816, USA }
\email[O. Atabek: ]{osman.atabek@u-psud.fr}

\date{\today}

\begin{abstract}
With a specific choice of laser parameters resulting into a so-called exceptional point in the wavelength-intensity plane, it is possible to produce the coalescence of two Floquet resonances describing the photodissociation of the molecule Na$_2$, which is one of the candidates for molecular cooling. Appropriately tuning laser parameters, following a contour around the exceptional point, the resonances exchange their labels. This represents a laser control of the vibrational transfer from one field-free state to another, through an adiabatic transport involving these resonances. The proportion of undissociated molecules at the end of the pulse is checked through Floquet adiabatic theory. A vibrational cooling scenario can be proposed based on a complete vibrational transfer which is predicted, with only 20 percent of molecules undergoing dissociation.
\end{abstract}


\maketitle


Progress made during the last two  decades in the cooling of quantum gases at ultracold temperatures down to the quantum degenerate regime has opened a possibility to study small molecules at the level of individual quantum rovibrational and hyperfine states \cite{ospelkaus2009,danzl2010}. An important step of the experimental developments is to achieve the deterministic manipulation of quantum states of small molecules, the formation of pre-selected quantum states and the control of the coupling between such states. The variety of robust experimental techniques \cite{dulieu2009,carr2009} which are currently being developed may later be used as a tool set for many-body physics, high precision measurements and technological applications such as quantum computing. Ultracold alkali-metal molecules are most often prepared in a distribution of vibrational levels of the singlet electronic ground state, or of the lowest triplet state. An important challenge is to create dense samples of ultracold molecules in a selected single quantum state. At present, there are many different techniques used to create and manipulate well-defined rovibrational and/or hyperfine states of diatomic molecules at ultralow temperatures, \textit{e.g.} photoassociation (PA) \cite{jones2006}, stimulated Raman adiabatic passage (STIRAP) \cite{guerin2003,danzl2008,lang2008,ni2008}, black-body radiation-assisted laser cooling \cite{staanum2010,schneider2010}, or optical pumping by a broadband femtosecond laser \cite{viteau2008,sofikitis2010}.

In this work we  propose an alternative way for a deterministic manipulation of vibrational states of alkali-metal diatomic molecules, relying on adiabatic transfer of population using chirped picosecond laser pulses, blue-detuned with respect to an atomic transition frequency. The technique is based on the coalescence of two resonance energies described using the Floquet formalism \cite{atabek2003}. The laser-controlled vibrational transfer scenario consists in applying an electromagnetic field which couples the ground or a metastable electronic molecular state to an excited electronic state. The induced resonances are identified through a series of complex eigenenergies of the Hamiltonian of the system, the imaginary part being related to their dissociation lifetime. With an appropriate choice of the laser parameters (intensity, wavelength) which is called an exceptional point (EP) \cite{kato1995,heiss1999} of the parameter space, it is possible \cite{lefebvre2009} to produce a degeneracy of two such energies. A specific loop encircling the EP in the laser parameter plane has to be identified, such that starting from a given field-free vibrational state,  a single resonance  adjusts continuously its characteristics (energy and width) to reach a different field-free vibrational state \cite{hernandez2006,heiss2004}.  In a second step we introduce the time requested to follow the loop and fulfill the conditions for an adiabatic transfer. This finally leads to an estimate of the percentage of undissociated molecules at the end of the pulse.

The proposed scheme is general but we consider hereafter the example of the Na$_2$ molecule in an excited vibrational level of its a$^3\Sigma_u^+$ lowest triplet electronic state (referred to as state $u$), which can be formed in optical traps \cite{shaffer2001}. The formation of ultracold sodium molecules in particular, has been demonstrated in several experiments \cite{lett1993,fatemi2002,amelink2003}, and therefore the present application represents a realistic situation. The same technique can be applied to other alkali dimers formed in optical traps with only small changes due to the similarities  of molecular potentials for all alkali dimers. The laser-controlled vibrational transfer scenario consists in applying an electromagnetic field with wavelength around 560~nm that couples state $u$ with the (1)$^3\Pi_g(3^{2}S+3^{2}P)$ excited electronic state, labeled $g$ below. As a complement to our preliminary study on H$_2^+$ (X$~^2\Sigma_g^+$ $\rightarrow$ A$~^2\Sigma_u^+$ photodissociation), the present situation looks quite promising. The reduced mass is about 20 times larger and the potential well of the initial state is about 200 times shallower, than in the H$_2^+$ case, so that the density of vibrational levels is about 2 orders of magnitude larger than in H$_2^+$. We thus expect resonance coalescence to occur at significantly smaller intensities, while the proximity of levels will require larger pulse times to remain in the adiabatic regime. Ultimately, we show that about 80\% of the Na$_2$ molecules survive to photodissociation, in contrast with the 10\% amount found for H$_2^+$.

\begin{figure}
\begin{center}
\includegraphics[width=0.45\textwidth]{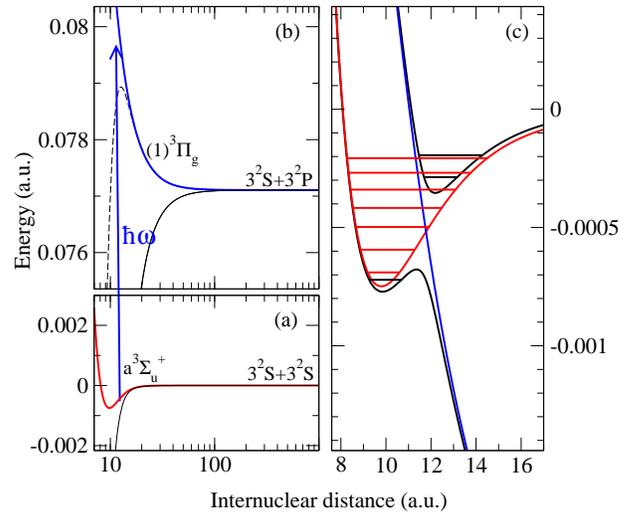}
\caption{\label{fig:1}
(Color online) The Na$_2$ potential curves involved in the proposed scheme. (a) the a$^3\Sigma_u^+(3^{2}S+3^{2}S)$, and (b) the (1) $^3\Pi_g(3^{2}S+3^{2}P)$ potentials. The absorption of a photon with energy $\hbar\omega=0.0808$~a.u. is pictured with a vertical arrow. (c): the same potentials dressed by the laser field to produce ($g$, -1) and ($u$, 0) channels. The adiabatic potentials (black lines) obtained by a diagonalization of the potential matrix including the laser-induced coupling are also shown. Horizontal lines indicate positions of bound and quasi-bound levels obtained after solving the Schr\"odinger equation for the corresponding one-channel adiabatic (black) or diabatic (red) potential curves.}
\end{center}
\end{figure}

Hereafter we refer to a model describing a rotationless field-aligned molecule in one spatial dimension (the internuclear distance $R$) and involving only two electronic states $\arrowvert u \rangle$ and $\arrowvert g \rangle$ whose Born-Oppenheimer (BO) potential energy curves $V_u^{BO}(R)$ and $V_g^{BO}(R)$ are displayed in Fig. \ref{fig:1}a, b. Among all electronic states of Na$_2$, the chosen $\arrowvert u \rangle$ and $\arrowvert g \rangle$ states present a favorable curve crossing situation at the laser wavelengths used to couple them (Fig. \ref{fig:1}c). In the following we neglect the spin-orbit interaction in the Na$_2$ molecule as it does not change the overall picture. The frozen rotation assumption can be validated by considering the long rotational periods of Na$_2$ (estimated as hundreds of ps) as compared with the pulse durations under consideration (less than 2 ps). The time-dependent wave function of the system is expanded on these two states
\begin{equation}
\label{eq:1}
 \arrowvert \Psi (R,t) \rangle = \chi_u(R,t)\arrowvert u \rangle +
 \chi_g (R,t) \arrowvert g \rangle\,,
\end{equation}
with the unknown functions $\chi_u$, $\chi_g$  accounting for the field-assisted nuclear dynamics.
When the interaction of the molecule with a continuous wave (cw) field with pulsation $\omega$ is described by a time-periodic Hamiltonian, applying the Floquet ansatz, the two-component solution of the time-dependent Schr\"odinger equation (TDSE) can be written as
\begin{eqnarray}
\left[ \begin{array}{c} \chi_u (R,t) \\
\chi_g (R,t) \end{array}\right] = e^{-iE_Ft/\hbar} \left[ \begin{array}{c} \phi_u (R,t) \\
\phi_g (R,t) \end{array}\right]\,.
\end{eqnarray}
where $E_F$ is the Floquet quasi-energy. The periodicity in time of $\phi_k(R,t)$ ($k=u,g$) allows us to expand these functions in Fourier series
\begin{equation}
\phi_k(R,t)=\sum_{n=-\infty}^{+\infty} e^{in\omega t}\varphi_{k}^{n}(R)\,.
\end{equation}
Finally, adopting the length-gauge and the long-wavelength approximation for the matter-field coupling, the Fourier components $\varphi_{k}^{n}(R)$ are given as solutions of the set of coupled differential equations resulting from the TDSE, exemplified by the general form, for all $n$ \cite{atabek2003}
\begin{equation}
\Big[T +V_{u,g}^{BO}+n\hbar\omega-E_F\Big]\varphi_{u,g}^{n}-1/2~ \mathcal{E}_0~ \mu(R)\Big[
\varphi_{g,u}^{n-1} + \varphi_{g,u}^{n+1}\Big]=0,
\end{equation}
\noindent where $T=(-\hbar^2/(2\,\mathcal{M}))(d^2\hspace*{-1.00mm}/dR^2)$ is the usual radial kinetic energy operator and $\mathcal{M}$ is the reduced mass. The transition between the $u$ and $g$ states is governed by the $R$-dependent molecular transition dipole moment $\mu(R)$ \cite{aymar_pc,aymar2005}. The laser is characterized by the linearly-polarized electric field $\mathcal{E}(t)=\mathcal{E}_0\cos(\omega t)$, the wavelength $\lambda= 2 \pi c /\omega$ and the intensity $I\propto \mathcal{E}_0^{2}$. Solutions with Siegert outgoing wave boundary conditions in the open channels produce  complex quasienergies of the form $E_F=E_R-i\Gamma_R/2$ where $\Gamma_R$ is the resonance width related to its decay rate.  The single photon processes are described using the two diabatic channels $u$ and $g$, with field-dressed potentials $V_u(R) \equiv V_u^{BO}(R)$, $V_g(R) \equiv V_g^{BO}(R)-\hbar \omega$ (Fig.\ref{fig:1}c), and with wave functions $\varphi_u^{0}(R)$ and $\varphi_g^{-1}(R)$, designated briefly as $\varphi_u(R)$ and $\varphi_g(R)$.  It is of course possible, for each value of the coordinate $R$, to obtain, through a linear transformation, the functions associated with the adiabatic potentials $V_{-}(R)$, $V_{+}(R)$ (Fig.\ref{fig:1}c) resulting from the diagonalization of the diabatic potential matrix after accounting for the matter-field interaction (wavelength $\lambda=563.903$~nm and intensity $I=0.35$~GW/cm$^2$) \cite{chrysos1993}.

The adiabatic potentials are very useful for the interpretation of resonances. In particular, for high laser intensities we expect the states associated with these adiabatic potentials to be very good zeroth-order approximations for the resonances of the full problem. We have previously observed \cite{atabek2008} that the Floquet resonances induced by high intensity electromagnetic fields fall into two categories: (i) the {\it Feshbach-type} (FT) resonances associated with the levels of the $V_{+}(R)$ potential well interacting with the continuum of the $V_{-}(R)$ potential; (ii) the {\it shape-type} (ST) resonances, decaying through or above the barrier of $V_{-}(R)$ and still interacting with $V_{+}(R)$ by residual non-adiabatic kinetic couplings. The energy of FT resonances increases with increasing intensity, as the adiabatic potentials split further and further; these resonances ultimately merge with the vibrational levels of $V_{+}(R)$ of vanishing width for very high intensity. Similarly the energy of ST resonances decreases with increasing intensity as the barrier height of the $V_{-}(R)$ potential decreases. Their widths grow until the flattened barrier is no more able to accommodate a resonance.

\begin{figure}
\begin{center}
\includegraphics[width=0.50\textwidth]{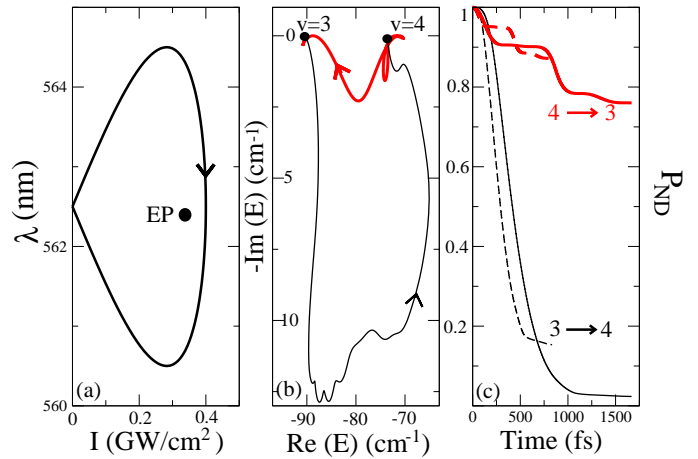}
\caption{\label{fig:2}
(Color online) (a) The laser pulse in the parameter plane ($\lambda$, $I$) producing (b) the trajectories of the energy plane either starting from the field-free vibrational level $v=3$ (black dashed line) or  $v=4$ (red solid line). Because the loop goes around the EP, there is a resonance exchange.
(c) The probability $P_{ND}$ for the molecule to survive after being exposed to 2 chirped pulses of duration 835 (dashed lines) and 1670 femtoseconds (solid lines). Two cases are considered: transitions  from $v=3$ to $v=4$ (black lines) and  from $v=4$ to $v=3$ (red lines).
 }
\end{center}
\end{figure}

In order to locate the EP resulting from the coalescence of FT and ST resonances, we calculate the resonance energies as a function of intensity for various values of the wavelength \cite{hernandez2006, lefebvre2009}.
Our best estimate for the parameters leading to coalescence of the $v=3, 4$ pair is  $\lambda_{EP}=562.53$~nm and $I_{EP}=0.332$~GW/cm$^{2}$. In conformity with our previous analysis \cite{lefebvre2009},  for $\lambda \le \lambda_{EP}$ the widths as a function of intensity, exhibit an avoided crossing with a tweezer-like profile, while the energies cross each other at the EP. These patterns are exchanged between widths and energies for $\lambda \ge \lambda_{EP}$. This means also that with a small change of the wavelength (a slightly larger value in our case) we can produce an avoided crossing for the energies while the rates will be crossing.
According to \cite{hernandez2006}, it should be possible to find a contour in the laser parameter plane that goes continuously from one resonance to another around an EP. A workable control scheme would be to apply  adiabatically a chirped pulse with the goal to transfer population from one field-free state to another. Adiabaticity refers here to a strategy consisting in the transport of a vibrational level on a single Floquet resonance, following it in time until it changes its label and ends up into another single vibrational level. This requires a laser pulse duration $t_f$ that exceeds the vibrational transition time, roughly estimated as $\hbar/\Delta E_v$, $\Delta E_v$ being the energy gap between adjacent vibrational levels. The resulting penalty, when taking a long duration laser pulse, is a possible depletion of molecular bound states. It is thus very important to reach a compromise between adiabaticity and the requirement for a dissociation not to be complete along the contour. We have built such a contour with an intensity and a wavelength obeying to a simple relation
\begin{equation}
I=I_{max}~\sin(\phi/2),~~~~~~\lambda=\lambda_{0}+\delta \lambda~ \sin(\phi)\,,
\end{equation}
with $\lambda_{0}=562.53$ nm, {\it i.e.} the wavelength for the EP of the pair $3, 4$, $I_{max}=0.4$~GW/cm$^{2}$ and $\delta \lambda=2$ nm. The parameter $\phi$ is continuously and linearly changing with time from $0$ (at $t=0$) to $2 \pi$ (at $t=t_f$). It is important to emphasize the experimental feasibility of such a picosecond laser pulse in the visible domain, with a chirped amplitude around 560 nm not exceeding $1\%$ and a rather modest intensity, less than 0.5 GW/cm$^2$. Figure \ref{fig:2} shows the loop in the parameter plane (panel a) and the resulting trajectories in the energy plane (panel b). The trajectories going from $v=3$ to $v=4$ or from $v=4$ to $v=3$, involve different rates, in agreement with the previous analysis of ST and FT resonances \cite{lefebvre2009}.

In order to estimate the fraction of undissociated molecules left after the pulse is over, we refer to the adiabatic Floquet theory \cite{atabek2008,fleischer2005}. This assumes that the chirped laser pulse envelope and its frequency vary sufficiently slowly with time such that the overall fraction of nondissociated molecules $P_{\rm ND}$  is given by
\begin{equation} \label{P-undiss}
   P_{\rm ND}(t) \; = \; \exp\left[- \hbar^{-1}\int_0^{t} \Gamma_R(t')~dt'\right] \;\; .
\end{equation}
Here, $\Gamma_R(t)$ is associated with the relevant Floquet quasi-energy eigenvalue calculated using the instantaneous
field parameters at time $t$. We present in Fig. \ref{fig:2}c $P_{\rm ND}$ for two cases: either starting from $v=3$ to reach $v=4$, or the reverse process and for two different laser durations $t_f$. The results differ drastically. As already mentioned, the interpretation relies on the nature of the instantaneous resonance states created during the pulse. In the first case ($v=3$ to $v=4$), during the first half of the pulse with increasing intensity and $\lambda \geq \lambda_{EP}$ the resonance is ST, with large values of the decay rates. In the second half of the pulse with decreasing intensity it also matches with the ST resonance $v=3$. For the transfer $v=4$ to $v=3$ the situation is opposite: The FT resonance of the first half of the pulse matches with the FT resonance of the second half, after the exchange of label when going across $\lambda_{EP}$. Therefore there is a better chance to preserve the molecules from dissociating when the transfer amounts to a decrease of the vibrational quantum number.

A crucial result of this work is that at the end of the pulse more than $80\%$ of the molecules survive the dissociation and are completely transferred from vibrational state $v=4$ to $v=3$. The importance of this result can be understood by comparing with a similar vibrational transfer control in the lighter molecular system H$_2^+$. As previously mentioned, due to much higher density of levels, resonance coalescences in Na$_2$ are obtained for much lower laser intensities. More specifically, $I_{EP} \simeq 0.3$ GW/cm$^2$ for Na$_2$, as compared with $I_{EP} \simeq 0.4$ TW/cm$^2$ for H$_2^+$. As a result the resonances participating in the adiabatic transfer process have decay rates of the order of ten cm$^{-1}$ while, in the case of H$_2^+$, they reach a thousand of cm$^{-1}$, leading thus to much better protection against dissociation for Na$_2$ . However, a high density of levels requires longer pulse durations, for the adiabaticity to be safely implemented. More precisely, 50 fs pulse  was long enough for an adiabatic transfer in H$_2^+$, whereas about 1000 fs  is necessary for the same adiabaticity to be achieved in Na$_2$, which is thus submitted to laser interaction and dissociation for a duration about 20 times  longer than H$_2^+$. Despite this fact, the long lifetimes of Na$_2$ Feshbach resonances on which the vibrational transfer relies are such that $80\%$ of Na$_2$ molecules remain non-dissociated at the end of the pulse, as compared to only $10\%$ in the case of H$_2^+$.
It is worth noticing that there are other lifetime-limiting processes competing with the photodissociation, namely spontaneous emission (for the (1)$^3\Pi_g$ state) and the dimer-atom inelastic collisions between Na and Na$_2$. The Na$_2$ lifetimes associated with these processes are much larger than 1000 fs, of the order of a few ns for the spontaneous emission and a few seconds for inelastic atom-dimer collisions.

We have shown that complete vibrational transfer is possible using resonance coalescence (exceptional points) in laser-induced photodissociation/photoassociation of a molecular system using a rather simple experimental technique. This is illustrated for the case of Na$_2$ with realistic laser parameters. Comparing the cases of H$_2^+$ and Na$_2$, we have shown that the proposed scheme can be applied to other molecular species like Cs$_2$ or Rb$_2$, because the wavelength and the amplitude of the applied electric field are two parameters  giving considerable freedom to manipulate the resonance quasienergies.
The ultimate goal for this and future similar studies is  molecular vibrational cooling using a purification scheme  to obtain a single ro-vibrational state, by successive transfers starting, for instance, from a thermal distribution of initial vibrational levels.

This work was done with the support of {\it Triangle de la Physique} (project 2008-007T-QCCM: Quantum Control of Cold Molecules), and of National Science Foundation under grant PHY-0855622.


%

\end{document}